\documentclass[pre,floatfix,superscriptaddress,notitlepage]{revtex4}
\pdfoutput=1
\usepackage{bbm, amsmath, graphicx, mathrsfs, float, mathtools, cases}
\usepackage{algorithm}
\usepackage{algorithmic}
\usepackage{hyperref}

\newcommand{\1}{\mathbbm 1}
\newcommand{\sign}{\mathrm{sign}}
\newcommand{\bx}{\mathbf x}
\newcommand{\bt}{\mathbf t}

\newcommand{\dd}{\partial}
\newcommand{\sm}{\setminus}

\begin{document}

\title{Optimizing spread dynamics on graphs by message passing}

\author{F.~Altarelli}
\affiliation{DISAT and Center for Computational Sciences, Politecnico di Torino, Corso Duca degli Abruzzi 24, 10129 Torino, Italy}
\affiliation{Collegio Carlo Alberto, Via Real Collegio 30, 10024 Moncalieri, Italy}
\author{A.~Braunstein}
\affiliation{DISAT and Center for Computational Sciences, Politecnico di Torino, Corso Duca degli Abruzzi 24, 10129 Torino, Italy}
\affiliation{Human Genetics Foundation, Via Nizza 52, 10126 Torino, Italy}
\affiliation{Collegio Carlo Alberto, Via Real Collegio 30, 10024 Moncalieri, Italy}
\author{L.~Dall'Asta}
\affiliation{DISAT and Center for Computational Sciences, Politecnico di Torino, Corso Duca degli Abruzzi 24, 10129 Torino, Italy}
\affiliation{Collegio Carlo Alberto, Via Real Collegio 30, 10024 Moncalieri, Italy}
\author{R.~Zecchina}
\affiliation{DISAT and Center for Computational Sciences, Politecnico di Torino, Corso Duca degli Abruzzi 24, 10129 Torino, Italy}
\affiliation{Human Genetics Foundation, Via Nizza 52, 10126 Torino, Italy}
\affiliation{Collegio Carlo Alberto, Via Real Collegio 30, 10024 Moncalieri, Italy}

\begin{abstract}
Cascade processes are responsible for many important phenomena in natural and social sciences.
Simple models of irreversible dynamics on graphs, in which nodes activate depending on the state of their neighbors, have been succesfully applied to describe cascades in a large variety of contexts.
Over the last decades, many efforts have been devoted to understand the typical behaviour of the cascades arising from initial conditions extracted at random from some given ensemble. However, the problem of optimizing the trajectory of the system, i.e. of identifying appropriate initial conditions to maximize (or minimize) the final number of active nodes, is still considered to be practically intractable, with the only exception of models that satisfy a sort of {\em diminishing returns} property called submodularity.
Submodular models can be approximately solved by means of greedy strategies, but by definition they lack cooperative characteristics which are fundamental in many real systems.
Here we introduce an efficient algorithm based on statistical physics for the optimization of trajectories in cascade processes on graphs. We show that for a wide class of irreversible dynamics, even in the absence of submodularity, the spread optimization problem can be solved efficiently on large networks. Analytic and algorithmic results on random graphs are complemented by the solution of the spread maximization problem on a real-world network (the Epinions consumer reviews network).
\end{abstract}

\maketitle
\tableofcontents

\section{Introduction}
Recent progress in the knowledge of the structure and dynamics of complex natural and technological systems has led to new challenges on the development of algorithmic methods to control and optimize their behavior.
This is particularly difficult for large-scale biological or technological systems, because the intrinsic complexity of the control problem is magnified by the huge number of degrees of freedom and disordered interaction patterns. Even simple to state questions in this field can lead to algorithmic problems that are computationally hard to solve.
For instance, the problem of identifying a set of initial conditions which can drive a system to a given state at some later time is in most cases computationally intractable, even for simple irreversible dynamical processes.
This is however a very relevant problem because irreversible spreading processes on networks have been studied since long time in several fields of research, ranging from percolation in statistical physics \cite{CLR79}, the spread of innovations and viral marketing \cite{G78,R83,JY05,AOY11}, epidemic outbreaks and cascade failures \cite{NE02,W02}, avalanches of spiking neurons \cite{OCK13}, and financial distress propagation in the interbank lending networks \cite{financial-risk,Nier2007,Gai2010,HM11}.

In typical applications, the nodes of a network can be in one of two states (e.g. ``active'' and ``inactive''), according to some dynamical rule which depends on the states of neighboring nodes.
Computer simulations and mean-field methods have shed light on the relation between the average behavior of macroscopic quantities of interest,
such as the concentration of ÒactiveÓ nodes, and the topological properties of the underlying network (e.g. degree distribution, diameter, clustering coefficient),
but they provide no effective instrument to optimally control these dynamical processes \cite{BBV08}.
Overcoming this limitation is crucial for real-world applications, for instance in the design of viral marketing campaigns, whose goal is that of targeting a set of
individuals that gives rise to a coordinated propagation of the advertisement throughout a social network achieving the maximum spread at the minimum cost. A similar optimization process can be applied to identify which sets of nodes are
most sensitive for the propagation of failures in power grids or which sets of banks are ``too contagious to fail'', providing a tool to assess and prevent systemic risk in large complex systems.
It is known that simple heuristic seeding strategies, such as topology-based ones, can be used to improve the spread of the process \cite{K10} as compared to random seeding. In order to attack this optimization problem in a systematic way, one should be able to analyze the large deviations properties of the dynamical process, i.e. exponentially rare dynamical trajectories corresponding to macroscopic behaviors that considerably deviate from the average one. At the ensemble level, this was recently put forward in \cite{nostro}. In the present work we consider the optimization aspects, developing a new algorithm based on statistical physics that can be applied to efficiently solve
this type of {\em spread optimization problems} on large graphs. As a case study, we provide a validation and a comparative analysis of the maximization of the spread of influence in a large scale social network (the ``Epinions'' network). Our results show that the performance of the optimization depends on the nature of the collective dynamics and, in particular, on the presence of cooperative effects.

The paper is organized as follows. In Section \ref{sec-sop} we introduce the spread optimization problem, focusing on a particular model of irreversible propagations on graphs, and we discuss the related literature.
Section \ref{sec-mp} is devoted to the derivation of the message-passing approach to the spread optimization problem, while the validation of the method on computer-generated graphs and the results obtained on the real-world network are presented in Section \ref{sec-result}.

\section{The Spread Optimization Problem}\label{sec-sop}
\subsection{Related Work}
The problem of maximizing the spread of a diffusion process on a graph was first proposed by
Domingos and Richardson \cite{DR01} in the context of viral marketing on social networks. The problem can be summarized as follows:
given a social network structure and a diffusion dynamics (i.e. how the individuals influence each other), find the smallest (or less costly) set of influential  nodes such that by introducing them with a new technology/product,
the spread of the technology/product will be maximized.
Kempe et al. \cite{KKT03} gave a rigorous mathematical formulation to this optimization problem considering two representative diffusion models: the independent cascade (IC) model and the
Linear Threshold (LT) model. They studied the problem in a stochastic setting, i.e. when the dynamics is based on randomized propagation, showing that for any fixed parameter $k$ it is NP-hard to find the $k$-node set achieving the maximum expected spread of influence.
They proposed a hill-climbing greedy algorithm, analogous to the greedy set cover approximation algorithm \cite{CLRS01}, that is guaranteed to generate a solution with a final number of active nodes that is provably within a factor $(1-1/e)$ of the optimal value for both diffusion models \cite{KKT03}. In the greedy algorithm, the nodes that generate the largest expected propagation are added one by one to the seed set.  Because of the stochastic nature of the problem, the algorithm is computationally very expensive, because at each step of the process one has to resort to sampling techniques to compute the expected spread. Leskovec et al. \cite{L07} proposed a ``lazy-forward'' optimization approach in selecting new seeds that reduces considerably the number of influence spread evaluations, but it is still not scalable to large graphs. Chen et al. \cite{CWW} proved that the problem of exactly computing the expected influence given a seed set in the two models is \#P-hard and provided methods to improve the scalability of the greedy algorithm. For a more comprehensive review of techniques and results, see e.g. \cite{K07}.

Real diffusion processes on networks are likely to be intrinsically stochastic, and information on the properties of the real dynamics might be obtained by surveys or data mining techniques \cite{mining}.
Different forms of stochasticity seem to make great difference on the properties of the optimization problem. In fact, only a peculiar choice of the distributions of parameters (e.g. uniform thresholds on a given interval for the LT model \cite{KKT03}) makes possible the derivation of the approximation result and guarantees good performances of the greedy algorithm.
On the other hand, the apparently simpler case represented by the deterministic LT model \cite{G78} was proven to be computationally hard even to approximate in the worst case\cite{LZWKF}. This is due to the lack of a key property, known as submodularity \cite{NWF78}, that is necessary to prove the approximation bound of Kempe et al. \cite{KKT03} and that will be discussed later.

\subsection{The Linear Threshold model}

The deterministic Linear Threshold model was first proposed by Granovetter \cite{G78} as a stylized model of the adoption of innovation in a social group, motivated by the idea that an individual's adoption behavior is highly
correlated with the behavior of her contacts. The model assumes that the influence of individual $j$ on individual $i$ can be measured by a weight $w_{ji} \in \mathbbm R^+$, and that $i$'s decision on the adoption of an innovation only depends on whether or not the total influence of her peers that already adopted it exceeds some given personal threshold $\theta_i \in \mathbbm R^+$.
In a social network, represented by a directed graph $G = (V, E)$, individuals are the nodes of the graph and they can directly influence only their neighbors. At each time step, a node $i \in V$ can be in one of two possible states:  $x_i = 0$ called \emph{inactive} and $x_i = 1$ called \emph{active}, that corresponds to the case in which $i$ adopts the product.  A vertex which is active at time $t$ will remain active at all subsequent times, while a vertex which is inactive at time $t$ can become active at time $t+1$ if some threshold condition, depending on the state of its neighbors in $G$ at time $t$, is satisfied. More precisely, given an initial set of nodes that are already active at time $t=0$, the so-called {\em seeds} of the dynamics, the dynamical rule is as follows
\begin{equation}
 x_i^{t+1} =
 \begin{cases}
 1 & \text{if } x_i^t = 1 \text { or } \sum_{j \in \dd i} w_{ji} x_j^t \geq \theta_i \,, \\
 0 & \text{otherwise} \, ,
 \end{cases}
\end{equation}
where $\dd i$ is the set of the neighbors of $i$ in $G$.
Note that by interpreting empty sites as active nodes, the Bootstrap Percolation process from statistical physics \cite{CLR79} becomes a special case of the LT model.

\section{The Message-Passing Approach}\label{sec-mp}

\subsection{Mapping on a Constraint-Satisfaction Problem}
In the LT model, the trajectory $\{\bx^0, \dots, \bx^T\}$ (with $\bx^t = \{x_i^t, \, i \in V\}$) representing the time evolution of the system can be fully parametrized by a configuration $\bt = \{t_i, \, i \in V\}$, where $t_i \in \mathcal T = \{0, 1, 2, \dots, T, \infty \}$ is the activation time of node $i$ (we conventionally set $t_i = \infty$ if $i$ does not activate within an arbitrarily defined stopping time $T$) \footnote{In general, on each node one needs a discrete variable that labels all possible single-node trajectories. In a non-progressive model, the number of single-node trajectories grows exponentially with the maximum number of times a node is allowed to change state in $[0,T]$. This parametrization is thus computationally tractable only if such number is known a priori for every node and it is very small (of order $O(1)$).}.
Given a set of seeds $S=\{i: t_i=0\}$, the solution of the dynamics is fully determined for $i \notin S$ by a set of relations among the activation times of neighboring nodes, which we denote by $t_{i} = \phi_i(\{t_j\})$ with $j \in \dd i$.
In the LT model, the explicit form of the constraint between the activation times of neighboring nodes is then
\begin{equation}
t_i = \phi_i(\{t_j\}) = \min \left\{ t \in \mathcal T : \sum_{j\in\dd i} w_{ji}\1[t_j<t]\geq \theta_i \right\}
\end{equation}
where the indicator function $\mathbbm 1[\text{condition}]$ is 1 if `condition' is true and 0 otherwise. This constraint expresses the rule that the activation time of node $i$ is the smallest time at which the sum of the weights of its active neighbors reaches the threshold $\theta_i$.
Admissible dynamical trajectories do correspond to configurations of activation times $\bt$ such that the binary function
$\Psi_i = \1[t_i = 0] + \1\left[t_i = \phi_i(\{t_j\})\right]$ equals $1$ for every node $i$.

We then introduce an energy function $\mathscr E(\bt)$ to give different probabilistic weights to trajectories with different activation times, $\mathbbm P[\bt] \propto \exp[-\beta \mathscr E(\bt)]$.
The specific form of the energy function will depend on the features of the trajectory that one wishes to select.
In the most general form, $\mathscr E(\bt) = \sum_i \mathscr E_i(t_i)$ where $\mathscr E_i(t_i)$ is the ``cost'' (if positive, or ``revenue'' if negative) incurred by activating vertex $i$ at time $t_i$.
When $\beta$ is large (compared to the inverse of the typical energy differences between trajectories), the distribution will be concentrated on those rare trajectories with an energy much smaller than the average, i.e. the large deviations of the distribution. When $\beta \to \infty$, only the trajectory (or trajectories) with the minimum energy are selected. Notice that the value chosen for $T$ will affect the ``speed'' of the propagation: a lower value of $T$ will restrict the optimization to ``faster'' trajectories, at the (possible) expense of the value of the energy.

Solving the Spread Maximization Problem (SMP) corresponds to selecting the trajectories that activate the largest number of nodes with the smallest number of seeds, or more precisely which minimize the energy function $\mathscr E(\bt) = \sum_i \left\{ c_i \1 \left[t_i = 0\right] - r_i \1 \left[ t_i < \infty \right] \right\}$ where $c_i$ is the cost of selecting vertex $i$ as a seed, and $r_i$ is the revenue generated by the activation of vertex $i$ (independently of the activation time). We consider the values of $\{c_i\}$ and $\{r_i\}$ as part of the problem definition, together with the graph $G$, the weights $\{w_{ij}\}$ and the thresholds $\{\theta_i\}$.
Trajectories with small energy will have a good trade-off between the total cost of the seeds and the total revenue of active nodes.

\subsection{Derivation of the Belief-Propagation equations}
The representation of the dynamics as a high dimensional static constraint-satisfaction model over discrete variables (i.e. the activation times) makes it possible to develop efficient message-passing algorithms, inspired by the statistical physics of disordered systems \cite{MP},  that can be used them find a solution to the SMP.

Our starting point is the finite temperature version ($\beta <\infty$) of the spread optimization problem, in which a Belief-Propagation (BP) algorithm can be used to analyze the large deviations properties of the dynamics  \cite{nostro}.

The Belief Propagation algorithm is a general technique to compute the marginals of locally factorized probability distributions under a correlation decay assumption \cite{MM09}. A {\em locally factorized distribution} for the variables $\bt = \{t_i, \, i \in V\}$ is a distribution which can be written in the form $P(\bt) \propto \exp[-F(\bt)]$ with $F(\bt) = \sum_a F_a(\bt_a)$ where each \emph{factor} contains only a (generally small) subvector $\bt_a$  of $\bt$. The {\em correlation decay assumption} means that in a modified distribution in which the term $F_a(\bt_a)$ is removed from the sum forming $F(\bt)$, the variables in $\bt_a$ become uncorrelated (the name \emph{cavity method} derives from the absence of this single term). The \emph{factor graph} representing the distribution is the bipartite graph where one set of nodes is associated to the variables $\{t_i\}$, the other set of nodes to the factors $\{F_a\}$ of the distribution, and an edge $(ia)$ is present whenever $t_i \in \bt_a$. When the factor graph is a tree, the correlation decay assumption is always exactly true. Otherwise a locally tree-like structure is usually sufficient for the decorrelation to be at least approximately verified.

\begin{figure}
\includegraphics[width=0.6\textwidth]{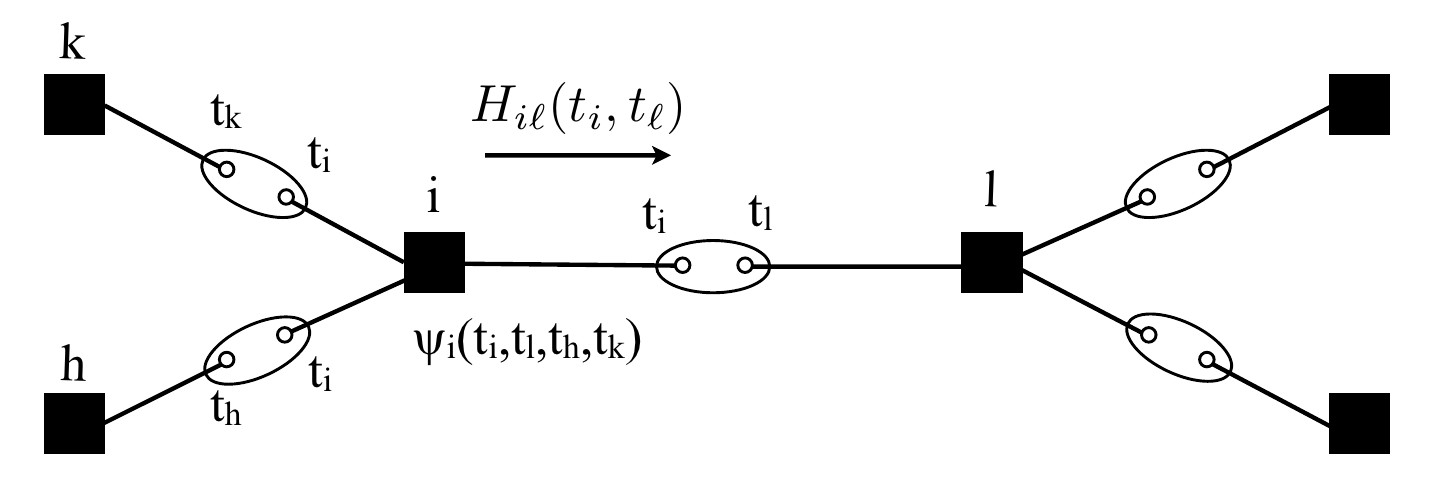}
\caption{Dual factor graph representation for the spread optimization problem}\label{fg}
\end{figure}

In the activation-times representation of the dynamics, the factor nodes $\{F_a\}$ are associated with the dynamical constraints $\Psi_i(t_i,\{t_j\}_{j\in \dd i})$ and with the energetic contributions $\mathscr{E}_i(t_i)$. Since nearby constraints for $i$ and $j$ share the two variables $t_i$ and $t_j$, it follows that the factor graph contains short loops. In order to eliminate these systematic short loops, we employ a dual factor graph in which variable nodes representing the pair of times $(t_i, t_j)$ are associated to edges $(i, j)\in E$, while the factor nodes are associated to the vertices $i$ of the original graph $G$ and enforce the hard constraints $\Psi_i$ corresponding as well as the contribution from $i$ to the energy function. Figure \ref{fg} gives an illustrative example of such dual construction.

The quantities that appear in the BP algorithm are called \emph{cavity marginals} or equivalently \emph{beliefs}, and they are associated to the (directed)  edges of the factor graph. We call $H_{i\ell}(t_i,t_\ell)$, the marginal distribution for a variable $(t_i,t_\ell)$ in the absence of the factor node $F_{\ell}$. Under the correlation decay assumption, the cavity marginals obey the equations \cite{nostro}
\begin{equation}\label{bp}
H_{ij}(t_i, t_{j}) \propto e^{-\beta\mathscr{E}_i(t_i)}\sum_{\{t_k\}_{k \in \dd i \sm j}}\Psi_i(t_i,\{t_k\}_{k \in \dd i})\prod_{k \in \dd i \sm j} H_{ki}(t_k , t_i).
\end{equation}
These self-consistent equations are solved by iteration. In Appendix \ref{ABP} we show how to compute these updates in a computationally efficient way not discussed in \cite{nostro}. Once the fixed point values of the cavity marginals are known, the ``full'' marginal of $t_i$ can be computed as
$P_i(t_i) \propto \prod_{j \in \dd i} H_{ji}(t_j, t_i)$ and the marginal probability that neighboring nodes $i$ and $j$ activate at times $t_i$ and $t_j$ is given by $P_{ij}(t_i, t_j)\propto H_{ij}(t_i,t_j) H_{ji}(t_j, t_i)$. In all equations the proportionality relation means that the expression has to be normalized.
Equation \eqref{bp} allows to access the statistics of atypical dynamical trajectories (e.g. entropies of trajectories or distribution of activation times). However it does not provide a direct method to explicitly find optimal configurations of seeds (in terms of energy).  To this end, we have to take the zero-temperature limit ($\beta \to \infty$) of the BP equations, and derive the so-called Max-Sum (MS) equations.

\subsection{Derivation of the Max-Sum equations}

Writing explicitly the constraints and the energetic terms,  the BP equations \eqref{bp} read
\begin{subnumcases}{\label{update1bp}  H_{i\ell}(t_i, t_\ell) \propto}
  e^{- \beta c_i} \sum_{\{t_j, \, j \in \dd i \sm \ell\}} \prod_{j \in \dd i \sm \ell} H_{ji}(t_j,0)& if $t_i = 0$, \label{update1bp t zero} \\
  \sum_{\substack{
	\{t_j, \, j \in \dd i \sm \ell\} \text{ s.t:} \\
	\sum_{k \in \dd i} w_{ki} \1 [t_k \leq t_i - 1] \geq \theta_i \, , \\
	\sum_{k \in \dd i} w_{ki} \1 [t_k < t_i - 1] < \theta_i
  }}  \prod_{j \in \dd i \sm \ell} H_{ji}(t_j,t_i)   & if $0 < t_i \leq T$, \label{update1bp t positive} \\
 e^{-\beta r_i } \sum_{\substack{
	\{t_j, \, j \in \dd i \sm \ell\} \text{ s.t:} \\
	\sum_{k \in \dd i} w_{ki} \1 [t_k < T] < \theta_i
  }}  \prod_{j \in \dd i \sm \ell} H_{ji}(t_j,\infty)  & if $t_i = \infty$. \label{update1bp t d}
\end{subnumcases}

We introduce the MS messages $h_{i\ell}(t_i, t_\ell)$ defined in terms of the BP messages $H_{i\ell}(t_i, t_\ell)$ as
\begin{align}
h_{i\ell}(t_i, t_\ell) = \lim_{\beta \to \infty} \frac 1 \beta \log H_{i\ell}(t_i, t_\ell).
\end{align}
Taking the $\beta \to \infty$ limit of the BP equations \eqref{bp}  we obtain
\begin{equation}
h_{i\ell}(t_i, t_\ell) =-\mathscr{E}_i(t_i) + \max_{\substack{\{t_j, j\in \dd i\sm \ell\} \text{ s.t:} \\ \Psi_i(t_i, \{t_j\})=1}}\quad \sum_{j\in \dd i\sm\ell}h_{ji}(t_j,t_i) + C_{i\ell} \label{eq:msfirst}
\end{equation}
and more explicitly
\begin{subnumcases}{\label{update1}  h_{i\ell}(t_i, t_\ell) =}
  \max_{\{t_j, \, j \in \dd i \sm \ell\}} \left[ \sum_{j \in \dd i \sm \ell} h_{ji}(t_j,0) \right] - c_i + C_{i\ell} & if $t_i = 0$, \label{update1 t zero} \\
  \max_{\substack{
	\{t_j, \, j \in \dd i \sm \ell\} \text{ s.t:} \\
	\sum_{k \in \dd i} w_{ki} \1 [t_k \leq t_i - 1] \geq \theta_i \, , \\
	\sum_{k \in \dd i} w_{ki} \1 [t_k < t_i - 1] < \theta_i
  }} \left[ \sum_{j \in \dd i \sm \ell} h_{ji}(t_j,t_i) \right] + C_{i \ell} & if $0 < t_i \leq T$, \label{update1 t positive} \\
  \max_{\substack{
	\{t_j, \, j \in \dd i \sm \ell\} \text{ s.t:} \\
	\sum_{k \in \dd i} w_{ki} \1 [t_k < T] < \theta_i
  }} \left[ \sum_{j \in \dd i \sm \ell} h_{ji}(t_j,\infty) \right] - r_i + C_{i \ell} & if $t_i = \infty$. \label{update1 t d}
\end{subnumcases}
where the additive constant $C_{i\ell}$ is such that $\max_{t_i, t_\ell} h_{i\ell}(t_i, t_\ell) = 0$.

For $t_i = 0$ the maximization is unconstrained, therefore it reduces to
\begin{align}
  h_{i\ell}(0,t_\ell) = \sum_{j \in \dd i \sm \ell} \left[ \max_{t_j} h_{ji}(t_j,0) \right] - c_i + C_{ij} \label{update t zero}.
\end{align}

In the other cases, the number of elementary operations needed to compute the updates \eqref{update1 t positive} and \eqref{update1 t d} is exponential in the
connectivity of the node being considered, making the implementation unfeasible even for moderate values of node degree. A convolution method and a simplification of both the update equations and the messages can be used to reduce them into a form which can be computed efficiently.

\subsection{Efficient computation of the MS updates}

To compute \eqref{update1 t positive} efficiently we start by noticing that the second constraint in the maximization can be disregarded. This can be done because, by dropping the second constraint, we allow for a delay between the time at which the threshold for vertex $i$ is exceeded and the time at which vertex $i$ activates (i.e. $t_i$). However, provided that all the costs $c_i$ and all the revenues $r_i$ are positive, for any given seed set the energy of the trajectory in the original problem is smaller than or equal to the energy of any trajectory compatible with the same seeds in the relaxed model (and any trajectory of the original problem is admissible for the relaxed one). Therefore the ground states of the two problems coincide.

We then introduce the functions
\begin{align}
  q_{1, \dots, r}(\theta, t) &= \max_{
	\substack{
	  \{ t_1, \dots, t_r \} \text{ s.t:} \\
	  \sum_{j = 1, \dots, r} w_j \1 \left[ t_j \leq t - 1 \right] = \theta_1
	}
  } \left[ \sum_{j = 1, \dots, r} h_{j}(t_j, t) \right]
\label{convolution q}
\end{align}
with the single index $q$'s
\begin{align}
  q_j(\theta, t) &= \max_{
	\substack{
	  t_j \text{ s.t:} \\
	  w_j \1[t_j \leq t-1] = \theta
	}
  } h_j(t_j, t) \\
  &=
	\begin{dcases}
	  \max_{t_j > t - 1} h_j(t_j, t) & \text{if $\theta = 0$,} \\
	  \max_{t_j \leq t - 1} h_j(t_j, t) & \text{if $\theta = w_j$,} \\
	  -\infty & \text{otherwise}
	\end{dcases} \label{convolution q1}
\end{align}
and with the convolution of two $q$'s given by
\begin{align}\label{q-convo}
	q_{1, \dots, r}(\theta, t) &= \max_{\theta', \theta'' \text{ s.t: } \theta' + \theta'' = \theta}
  \left\{ q_{1, \dots, r'}(\theta', t) + q_{r'+1, \dots, r}(\theta'', t) \right\} \,.
\end{align}
This convolution property allows to compute $q_{\dd i \sm \ell}$ in a time which is linear in the connectivity of node $i$.

From $q_{\dd i \sm \ell}(\theta, t)$ we can  compute
\begin{align}
  m_{i \ell} (\theta, t) &= \max_{
	\substack{
	  \{ t_1, \dots, t_r \} \text{ s.t:} \\
	  \sum_{j = 1, \dots, r} w_j \1 \left[ t_j \leq t - 1 \right] \geq \theta \,
	}
  } \left\{ \sum_{j = 1, \dots, r} h_{j}(t_j, t) \right\} \\
  &= \max_{ \theta' \text{ s.t: } \theta' \geq \theta}
  q_{\dd i \sm \ell}(\theta', t) \label{convolution m}
\end{align}
in terms of which \eqref{update1 t positive} gives:
\begin{align}
  h_{i\ell}(t_i, t_\ell) = m_{i \ell} \big(\theta_i - w_{\ell i} \1 \left[ t_\ell \leq t_i - 1 \right], t_i \big) + C_{i \ell} && \text{if $0 < t_i \leq T$} \,. \label{update t positive}
\end{align}

Similarly to compute \eqref{update1 t d} we introduce
\begin{align}
  s_{1, \dots, r}(\theta) &= \max_{
	\substack{
	  \{ t_1, \dots, t_r \} \text{ s.t: } \\
	  \sum_{j = 1, \dots, r} w_j \1 \left[ t_j < T \right] = \theta
	}
  } \left[ \sum_{j = 1, \dots, r} h_j (t_j, \infty) \right]
\label{convolution s}
\end{align}
with single index $s$'s
\begin{align}
  s_j(\theta) &=
	\begin{dcases}
	  \max \left\{ h_j(T, \infty) , h_j(\infty,\infty) \right\} & \text{if $\theta = 0$} \\
	  \max_{t_j < T} h_j(t_j, \infty) & \text{if $\theta = w_j$} \\
	  -\infty & \text{otherwise}
	\end{dcases} \label{convolution s1}
\end{align}
and convolution
\begin{align}
  s_{1, \dots, r}(\theta) &= \max_{\theta', \theta'' \text{ s.t: } \theta' + \theta'' = \theta} \left\{ s_{1, \dots, r'}(\theta') + s_{r'+1, \dots, r}(\theta'') \right\} \,.
\end{align}
Once $s_{\dd i \sm \ell}$ is computed, we can compute
\begin{align}
  u(\theta) &= \max_{
	\substack{
	  \{ t_1, \dots, t_r \} \text{ s.t: } \\
	  \sum_{j = 1, \dots, r} w_j \1 \left[ t_j < T \right] < \theta
	}
  } \left\{ \sum_{j = 1, \dots, r} h_j (t_j, \infty) \right\} \\
  &= \max_{\theta' \text{ s.t: } \theta' < \theta} s(\theta') \label{convolution P}
\end{align}
in terms of which \eqref{update1 t d} becomes:
\begin{align}
  h_{i \ell}(\infty, t_\ell) &= u \big( \theta_i - w_{\ell i} \1 \left[ t_\ell < T \right] \big) - r_i + C_{i \ell}. \label{update t d}
\end{align}

\subsection{Simplification of the MS messages}

A simplification of the messages follows from noticing that the dependence of $h_{i \ell}$ on $t_{i\ell}$ is almost trivial:
for fixed $t_i$, $h_{i \ell}$ is independent on $t_{\ell}$ if $t_i = 0$, it only depends on $\sign{\left(t_{\ell} - (t_i-1)\right)} \in \{-1,0,1\}$ if $0 < t_i \leq T$, and it
only depends on $\1[t_{\ell} <T]$ if $t_i = \infty$. We define a new set of messages $\tilde h_{i \ell}(t_i, \sigma)$ by
\begin{align}
  \tilde h_{i \ell}(t_i, \sigma) &=
	\begin{cases}
	  h_{i \ell}(t_i, t_i - 2) & \text{if $t_i > 1$ and $\sigma = 0$} \\
	  h_{i \ell}(t_i, t_i - 1) & \text{if $t_i > 0$ and $\sigma = 1$} \\
	  h_{i \ell}(t_i, t_i) & \text{if $\sigma = 2$} \\
	  -\infty & \text{otherwise}.
	\end{cases} \label{H tilde}
\end{align}
which can be inverted as
\begin{align}
  h_{i \ell}(t_i, t_\ell) &= \tilde h_{i \ell} \big( t_i, \hat \sigma(t_i, t_\ell) \big) \label{H vs H tilde} \\
  \hat \sigma(t_i, t_\ell) &= 1 + \sign \big( t_\ell - (t_i - 1) \big) \label{sigma hat}.
\end{align}
It is thus possible to reduce the number of messages per node from $O(T^2)$ to $O(T)$. In terms of the new messages, \eqref{convolution q}, \eqref{convolution q1}, \eqref{convolution s} and \eqref{convolution s1} become respectively:
\begin{align}
  q_{1,\dots, r}(\theta, t) &= \max_{t_r} \left\{ \tilde h_r \big( t_r, \hat \sigma(t_r, t) \big) + q_{1, \dots, r-1} \big( \theta - w_r \1 [ t_r \leq t - 1 ], t \big) \right\} \\
  q_j(\theta, t) &=
	\begin{dcases}
	  \max_{t_j > t - 1} \tilde h_j \big( t_j, \hat \sigma(t_j, t) \big) & \text{if $\theta = 0$} \\
	  \max_{t_j \leq t - 1} \tilde h_j(t_j, 2) & \text{if $\theta = w_j$} \\
	  -\infty & \text{otherwise}
	  \end{dcases} \\
  s_{1, \dots, r}(\theta) &= \max_{t_r} \left\{ \tilde h_r ( t_r, 2 ) + s_{1, \dots, r-1} \big( \theta - w_r \1 [ t_r < T ] \big) \right\} \\
  s_j(\theta) &=
	\begin{dcases}
	  \max \left\{ \tilde h_j(T-1,2), \tilde h_j(T, 2) \right\} & \text{if $\theta = 0$} \\
	  \max_{t_j < T-1} \tilde h_j( t_j, 2) & \text{if $\theta = w_j$} \\
	  -\infty & \text{otherwise}
	\end{dcases}
\end{align}
and \eqref{update t zero}, \eqref{update t positive} and \eqref{update t d} become:
\begin{subnumcases}{\label{eq:simplified} \tilde h_{i \ell}(t_i, \sigma) = }
  \sum_{j \in \dd i \sm \ell}\max_{t_j} \tilde h_{j i} \big( t_j, \hat \sigma(t_j, 0) \big) -  c_i + C_{i \ell} & if $t_i = 0$ and $\sigma = 2$, \label{eq:simplified-beg}
\\
  m(\theta_i - w_{\ell i}, t_i) + C_{i \ell} & if $1 < t_i \leq T$ and $\sigma = 0$, \\
  m(\theta_i - w_{\ell i}, t_i) + C_{i \ell} & if $0 < t_i \leq T$ and $\sigma = 1$, \\
  m(\theta_i, t_i) + C_{i \ell} & if $0 < t_i \leq T$ and $\sigma = 2$, \\
  u(\theta_i - w_{\ell i}) - r_i + C_{i \ell} & if $t_i = \infty$ and $\sigma = 0$, \\
  u(\theta_i) - r_i + C_{i \ell} & if $t_i = \infty$ and $\sigma > 0$, \\
  -\infty & otherwise.\label{eq:simplified-end}
\end{subnumcases}

\subsection{Time complexity of the updates}

The implementation of the MS updates \eqref{eq:simplified} would require $O(T k (k - 1) \theta_i^2)$ operations for a vertex of degree $k$, but a factor $k - 1$ can be saved by pre-computing the convolution $q_{1,...,i}$ and $q_{i,...,k}$ for each $i=1,...,k$ using $2 k$ convolution operations, and then computing the ``cavity'' $q_{1,...,i-1,i+1,...k}$ as a convolution of $q_{1,...,i-1}$ and $q_{i+1,...,k}$.

\subsection{Convergence and reinforcement}
\label{sec:reinf}
MS equations do not converge in some cases. In such situation, the {\em reinforcement} strategy described below is of help\cite{steiner}. The idea is to use the noisy information given by MS before convergence to slowly drive the system to a simpler one in which the equations do converge, hopefully without altering the true optimum too much. This can be achieved by adding an ``external field'' proportional to the total instantaneous local field, with the proportionality constant slowly increasing over time in a sort of ``bootstrapping'' procedure. In symbols, this corresponds with the following update equations between iterations $\tau$ and $\tau+1$:
\begin{eqnarray}
h^{\tau+1}_{i\ell}(t_i, t_\ell) &=&-\mathscr{E}_i(t_i) + \max_{\substack{\{t_j, j\in \dd i\sm \ell\} \text{ s.t:} \\ \Psi_i(t_i, \{t_j\})=1}}\quad \sum_{j\in \dd i\sm\ell}h^\tau_{ji}(t_j,t_i) + \lambda^\tau p^\tau_{i\ell}(t_i,t_\ell) + C_{i\ell}.\label{eq:msrein}\\
p^{\tau+1}_{i\ell}(t_i, t_\ell) &=& h^{\tau}_{i\ell}(t_i, t_\ell) + h^{\tau}_{\ell i}(t_\ell, t_i) + \lambda^\tau p^\tau_{i\ell}(t_i,t_\ell) + \tilde{C}_{i\ell}
\end{eqnarray}
where $\lambda^\tau = \gamma\tau$ for some $\gamma > 0$ (other increasing functions of $\tau$ give similar qualitative behaviours). The case $\gamma=0$ corresponds with the unmodified MS equations, and we observe that the number of iterations scale roughly as $\gamma^{-1}$, while the energy of the solution found increases with $\gamma$.

\section{Results}\label{sec-result}

\subsection{Validation of the algorithm on computer-generated graphs}

\begin{figure}[t]
\includegraphics[width=12.5cm]{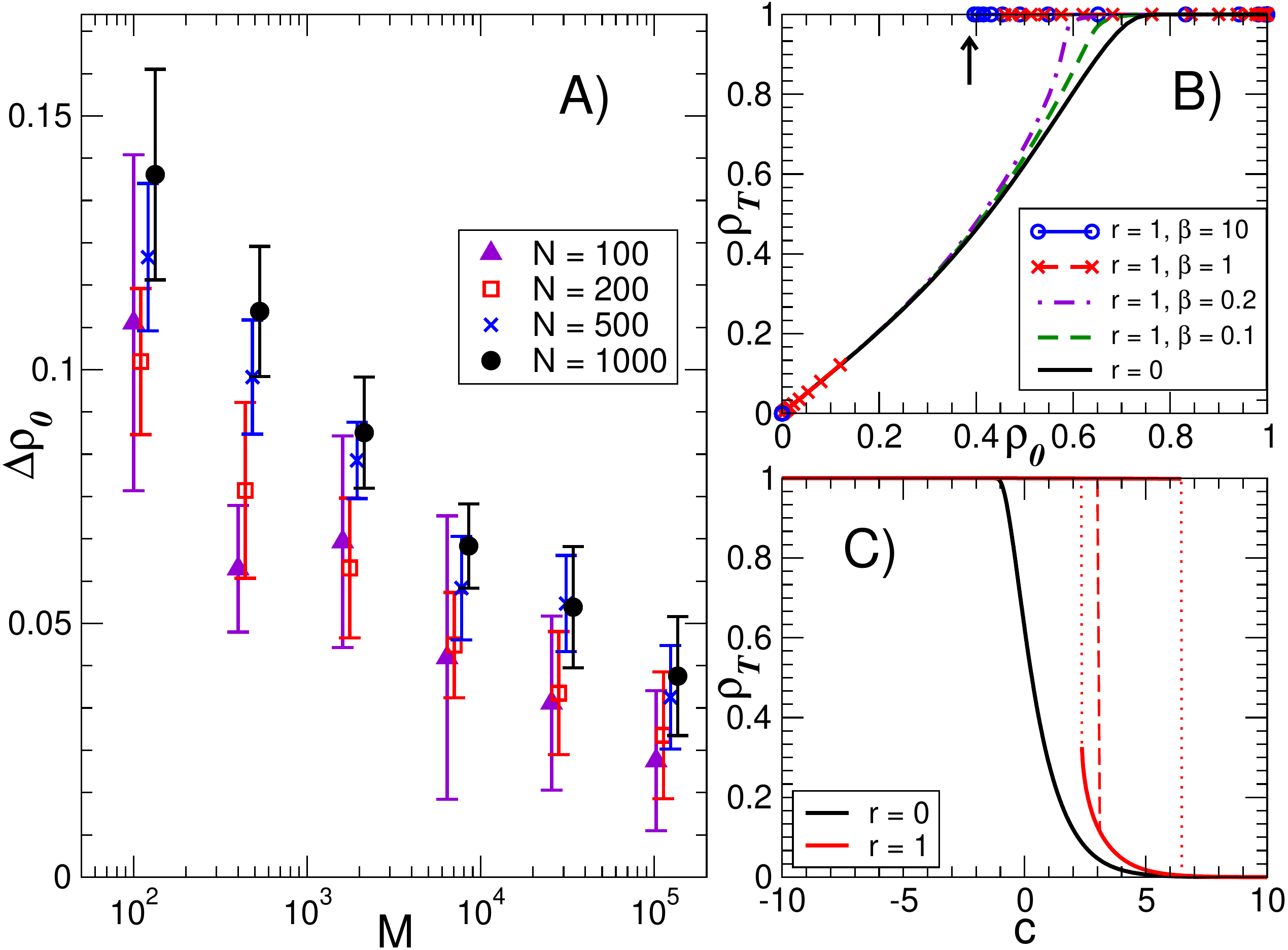}
\caption{A) Comparison between the performances of Simulated Annealing (SA) and Max-Sum (MS) algorithms on random regular graphs with degree $K=5$, threshold $\theta= 4$ and different sizes $N$. The quantity $\Delta\rho_0= \rho_0^{SA}/\rho_0^{MS} - 1$ is plotted for different system sizes $N$ as function of the total number of global updates $M$ employed in the SA. Points are slightly shifted horizontally from the correct value at $N=100$ for clarity.
B) Parametric plot $\rho_T$ v.s.$\rho_0$ obtained solving the BP equations in the single-link approximation on regular random graphs of degree $K=5$, for threshold $\theta = 4$, duration $T=20$ and $r = 1$ and different values of $\beta = 0.1$(green dashed line), 0.2 (violet dash-dotted line), 1 (red dashed line with crosses), and 10 (blue line with circles). The black line shows the solution for $r=0$. For $\beta = 1$ and $\beta = 10$ both $\rho_0$ and $\rho_T$ undergo a discontinuous jump as $c$ is varied. The vertical arrow indicates the minimum density of seeds ($\rho_0\approx 0.3862$) necessary for the total activation obtained by Max-Sum (MS) on finite graphs of size $N=10\,000$.
C) Curves $\rho_T(c)$ for $r = 0$ (black) and $1$ (red) with $\beta=1$. The red curves are obtained following the upper and lower branches of solution across the transition. Dotted vertical lines represent the stability limits of upper and lower solution branches. The red dashed line indicates the position of the thermodynamic transition.
}\label{figMC}
\end{figure}

We evaluated the performance of our method (both at finite $\beta$ and for $\beta \to \infty$) by studying the spread maximization problem of the LT model on regular random graphs with identical costs $c_i = c$ and revenues $r_i = r$ for all nodes. Though this choice might look peculiar, it is known that, in the average case, optimization problems can be very hard to solve even on completely homogeneous instances because of the onset of clustering (replica symmetry breaking) phenomena \cite{MP}.
Moreover, homogeneous instances provide a fair test-bed to compare different optimization methods, in which no additional source of topological information can be exploited to design ad-hoc seeding heuristics or to improve the performances of the algorithms.

When $r$ and $c$ are considered as free parameters (as opposed to having well defined values determined by the problem definition) the three parameters $r, c, \beta$ become redundant, and in the following we shall only consider the cases $r = 0$ with $c = 1$ and variable $\beta$, or $r=1$ with variable $c$ and $\beta$. Also, we shall consider as observables the density of seeds $\rho_0 = \frac{1}{N}\sum_{i} \1[t_i =0]$ and the final density of active nodes $\rho_T = \frac{1}{N}\sum_{i}\1[t_i < \infty]$ rather than the total cost and total revenue, since these would scale with $c$ and $r$ making it difficult to compare the results obtained for different values of $c$ and $r$.

In order to evaluate the results obtained with MS for $\beta \to \infty$, we considered three other strategies to minimize the energy $\mathscr{E}$: (1) Linear/Integer Programming (L/IP) approaches, (2) Simulated Annealing (SA) and (3) a Greedy Algorithm (GA). A detailed description of these algorithms is reported in Appendix \ref{Aalgo}.

The performances obtained by L/IP are extremely poor, becoming practically unfeasible for graphs over a few dozen nodes. For this reason, the results will not be reported in detail.
On the contrary the GA and SA algorithms gave results that can be directly compared with those obtained using the MS algorithm. The GA iteratively adds to the seed set the vertex that achieves the most favorable energy variation. It is relatively fast but it finds solutions with a number of seeds which is about 25\% larger than those obtained using MS. Simulated Annealing is considerably slower than GA and MS, and the running time necessary for SA to reach the MS solution scales poorly with the system's size. For a random regular graph of degree $K=5$ and thresholds $\theta =4$, we computed the quantity $\Delta\rho_0= \rho_0^{SA}/\rho_0^{MS} - 1$ (averaged over 10 realization of the graph) that represents the relative difference between the minimum density of seeds $\rho_0^{SA}$ required to activate the whole graph obtained using SA with exponential annealing schedule from $\beta = 0.5$ to $\beta = 10^3$ and the corresponding values $\rho_0^{MS}$ computed using the MS algorithm. The results for different system sizes $N$ are plotted in Figure \ref{figMC} as function of the total number of global updates $M$ employed in the SA (the total number of SA steps is $N M$). The decrease of $\Delta \rho_0$ with the number $M$ of Monte Carlo sweeps used to perform the annealing schedule from $\beta = 0.5$ to $\beta = 10^3$ is compatible with an exponentially slow convergence towards the results obtained using the MS algorithm. The running time of SA for instances with $N=1\,000, M=102\,400$ is several hours, while they are solved by MS in less than 1 minute on the same machine. Notice also that a relative difference of about 5\% in a solution is not small. For instance, on a graph of $N=1000$ nodes, the full-spread solution found using the MS algorithm counts $386$ seeds, while the best solutions found by SA have more than $400$ seeds. We did not report the GA results in Figure \ref{figMC} as they would be out of scale.

\begin{figure}[t]
\includegraphics[height=0.4\columnwidth]{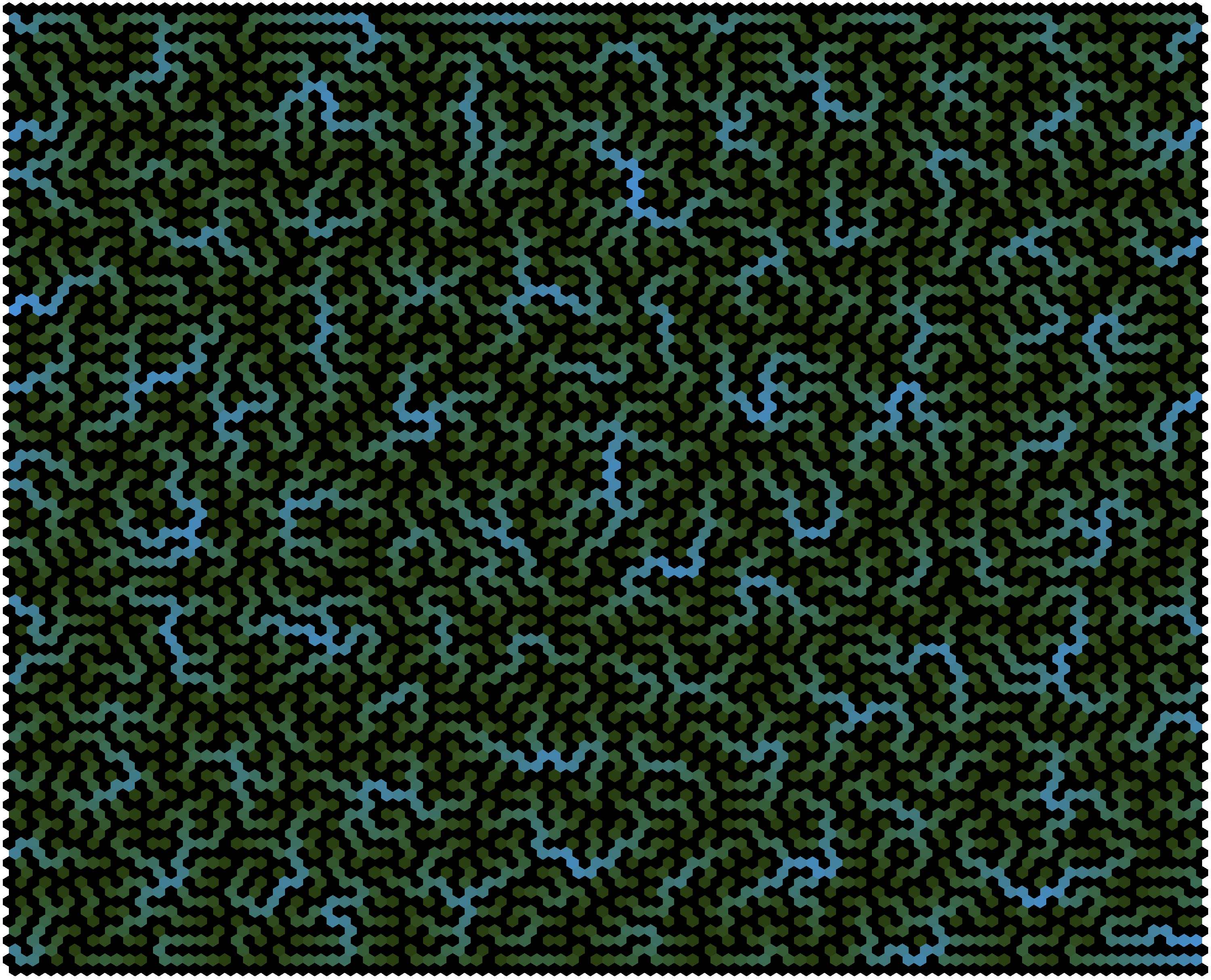}\hfill
\includegraphics[height=0.4\columnwidth]{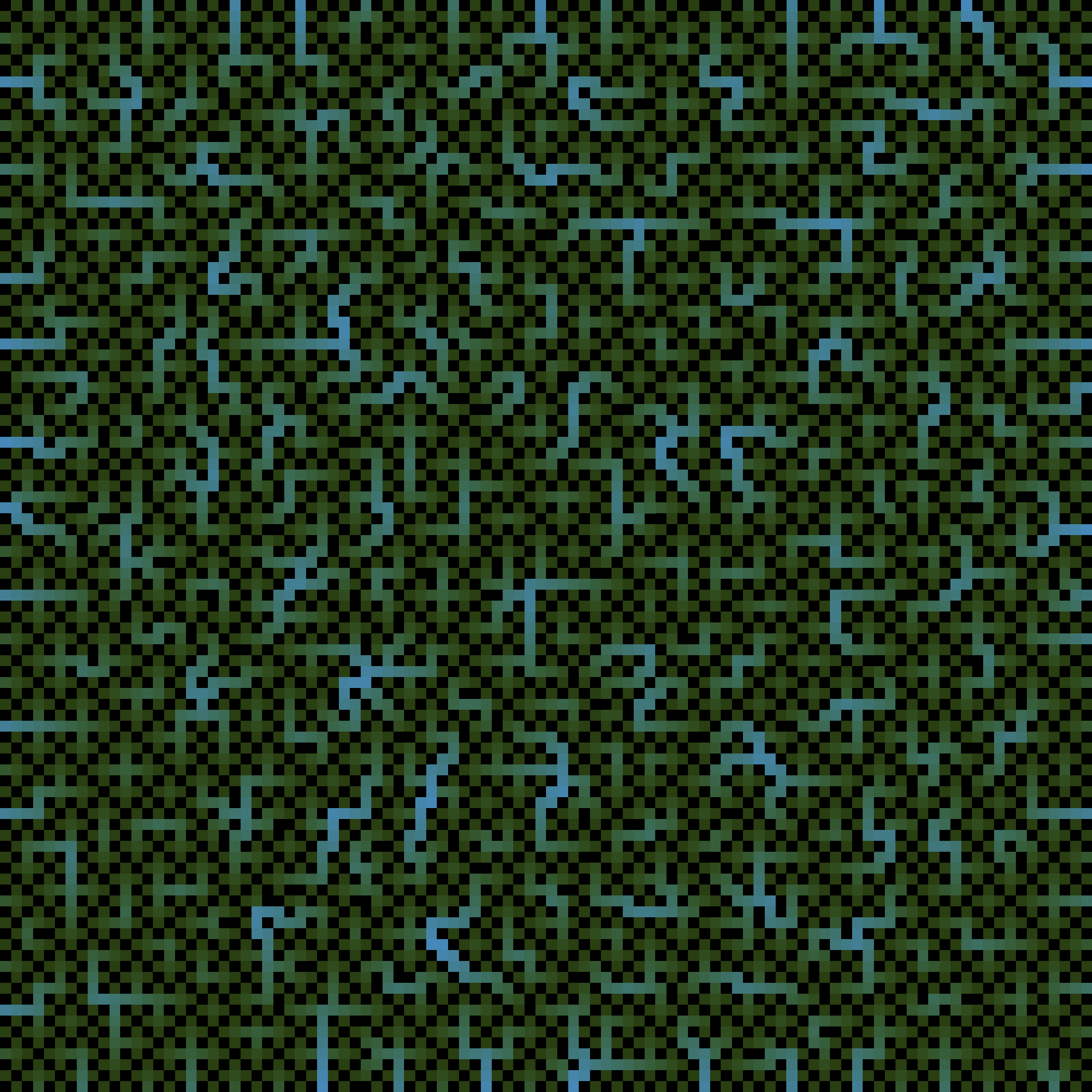}
\caption{Example of low seed density configurations found by MS equations with reinforcement in 2D $L\times L$ lattices with open boundaries. Colors correspond to activations times (from time zero for seeds in black to lighter colors). Left: hexagonal lattice ($K=6$) with $\theta=5, T=20, L=100$. Every hexagon is adjacent to at most one hexagon of a lighter color. A movie depicting the convergence of the reinforced MS equations is available as supplementary material. Right: rectangular lattice ($K=4$) with $\theta=3, T=15, L=100$
\label{fig:bacarozzi}
}
\end{figure}

On locally tree-like graphs such as large random regular graphs of finite connectivity the decorrelation assumption is approximately correct therefore we expect our MS algorithm to find solutions very close to the optimum. However
most real networks, in particular social networks, contain a non-negligible fraction of short loops. Even though our MS algorithm converges, in these cases there is no guarantee to find solutions close to optimality.
In order to investigate the performances of our algorithm on loopy graphs, we studied the spread maximization problem on some two-dimensional lattices.
Figure \ref{fig:bacarozzi} shows some examples of low seed density configurations obtained by MS on two-dimensional lattices with coordination numbers $K=6$ and $K=4$, thresholds $\theta = K-1$ and duration $T=20$ and $T=15$ respectively. In both cases we consider $L \times L$ lattices with $L=100$ and open boundary conditions which, due to the finite $T$, are not well suited for ordered solutions. Despite the fact that the cavity approximation is not expected to be asymptotically exact on lattices, the seed densities of the solutions we find, which are 0.5210 and 0.3536 respectively, are actually quite good. In fact, for the hexagonal lattice we can compare it with the seed density obtained by considering the $L \to \infty$ and $T \to \infty$ limits and taking a solution with the seeds arranged on parallel rows separated by one almost empty row, containing a single seed, which gives a seed density of $1/2$ (we stress that the corresponding solution obtained for $T=20$ and $L=100$ has a seed density of 0.5296, larger than the one obtained by MS). For the square lattice with $L\to \infty$ and $T \to \infty$ a self-similar solution with seed density $1/3$ can be constructed (and again, the corresponding solution with $T=15$ and $L \to \infty$ has seed density 0.3778, which is worse than what MS finds).

\subsection{Spread maximization in a social network}

As a test case, we studied the spread maximization problem for a large-scale real-world network.
We considered the network of trust relationships of the consumers' reviews website Epinions (\url{http://www.epinions.com}), one of the most studied in the context of spread optimization \cite{epinions}. Nodes represent users of the website and directed links $(i,j)$ indicate that user $i$ trusts user $j$. We used the LT model, setting $w_{ji} = 1$ if a directed link from $i$ to $j$ exists. We denote by $k_i^{in}$ (resp. $k_i^{out}$) the in-degree (resp. out-degree) of vertex $i$. Spread dynamics in such a network could represent the spreading of a consumer's choice, e.g. the adoption of a new product. For example, a user might adopt a new product if at least half of the users he trusts already use it. Correspondingly we set $\theta_i = \lfloor{(k_i^{out}+1)/2}\rfloor$.

This mechanism could be exploited for a viral marketing campaign aimed at promoting the new product. In this context, some investment needs to be done by the promoter company, in terms of goods for early adopters (either lower prices, free products or other forms of promotion). The goal is to maximize the total profit generated by the campaign. The price of the initial investment per early adopter and the revenue for an induced adopter are modeled by $c_i$ and $r_i$ respectively. We choose to consider a revenue $r_i = 1$ for all vertices  and cost $c_i=\mu (k^{in}_i + 1) + 1$ to represent the higher enrollment cost of highly influential individuals. For seeds, the extra 1 in $c_i$ cancels $r_i$, so that only ``activated'' nodes (i.e. active but not seeds) contribute positively to the revenue. This choice of $\theta_i$, $c_i$ and $r_i$, although clearly arbitrary, is reasonable in the context of viral marketing.

Figure \ref{figEpinions} displays the results obtained using MS, and four other algorithms (see Appendix \ref{Aalgo} for details): (1) Greedy algorithm based on energy computation (GA), (2) Greedy algorithm based on centrality using Kleinberg's HITS \cite{K99} procedure for directed networks (HITS), (3) Greedy algorithm based on in-degree (Hubs) and (4) Simulated Annealing (SA).
The performance of GA appears good for low values of the cost, but quickly deteriorates as the cost increases, reaching a plateau with a relatively small number of activated nodes.
With HITS we observe a sudden transition in the number of activated nodes as a function of the number of seeds, where it increases by an order of magnitude. With Hubs this transition occurs at a slightly smaller number of seeds.
Of course, due to the choice of the costs $c_i$, Hubs selects the most expensive seeds, which penalizes it, yet it outperforms GA and HITS in terms of energy. We tried several variants of the greedy algorithms in which seed preferences were altered by a linear function of the cost, but this did not improve performance in any case and thus will not be reported. The performance of SA depends strongly on two different choices: initial condition and annealing schedule. We obtained the best results using the solution found by Hubs as initial condition and an exponential annealing schedule starting from $\beta=0.01$ to $\beta=1000$ in around $10^7$ SA steps. This single run required around one week of simulation, but tuning the SA parameters required a much longer time. SA outperforms both GA and Hubs with these settings.
MS performance depends strongly on the time limit $T$. All values of $T$ lead to approximately the same number of activated nodes (which equals the revenue in this model), which is comparable with the one of SA, but they differ considerably in terms of cost. The case $T=15$ is close to the SA result, but for larger $T$ the cost decreases significantly while the number of activated nodes increases; the cost of SA is about $33\%$ higher than the one of MS for $T=40$.
If the propagation is allowed to continue for $T>40$ a small number of additional nodes activate.
It is worth noting that by varying $\mu$,  the behavior of the MS algorithm changes abruptly: the cost and the activation size jumps from (A) to a point close to (0,0) as a consequence of a lack of concavity of the cost-activation function. This phenomenon is the consequence of the existence of a discontinuous transition as function of $\mu$ that was first observed in a finite temperature analysis performed on random regular graphs with uniform thresholds   \cite{nostro}.

\begin{figure}[t]
\includegraphics[width=9.5cm]{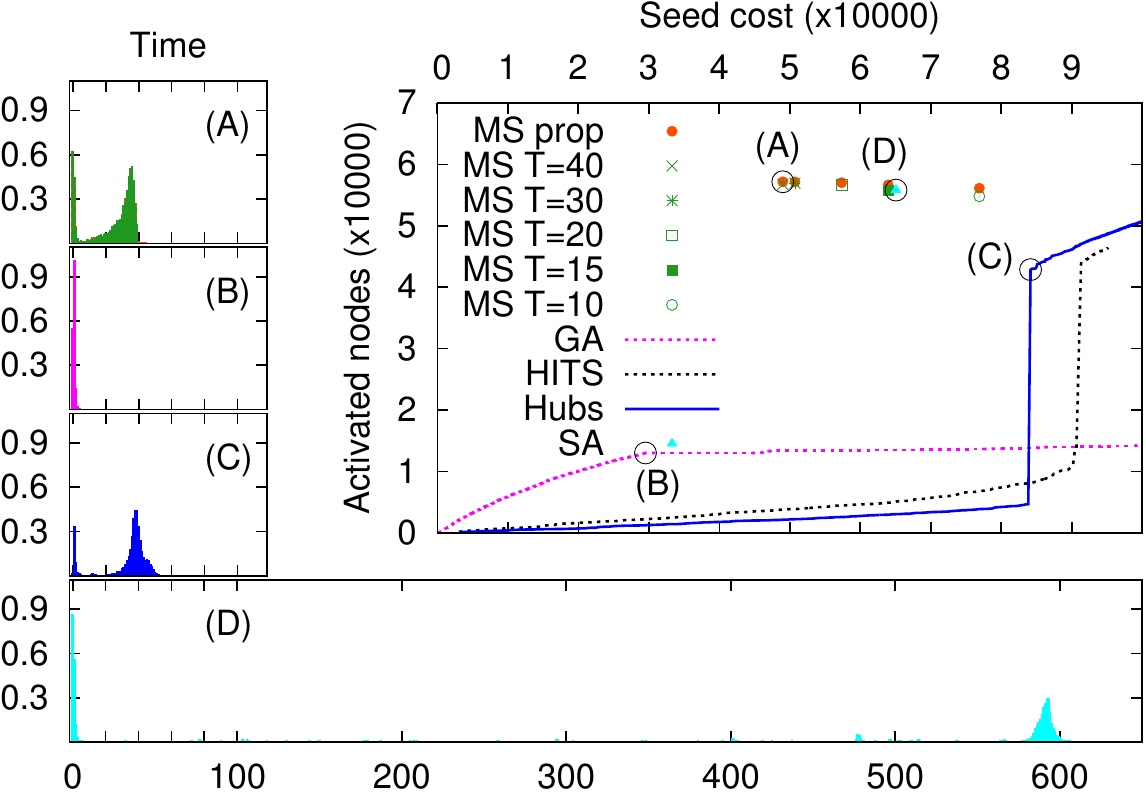}
\caption{
Optimization on the Epinions network with seed cost $c_i=\mu (k_i^{in}+1)+1$, with $k_i^{in}$ equal to the in-degree of node $i$ and $\mu=0.3$. Top Right: Number of activated nodes (i.e. active but not seeds) vs. total seed cost obtained by MS for values of $T$ from 10 to 40. As $T$ increases, the cost decreases and the number of propagated nodes slightly increases. Points labeled ``MS prop'' correspond to the unbounded time propagation of the same seeds obtained for different $T$'s; as $T$ increases, this extra propagation becomes irrelevant (as may be expected).  For comparison, performance of greedy based on energy computation (GA), greedy based on HITS procedure (HITS), the propagation obtained by the largest $n$ in-degree nodes (Hubs) and the best result of Simulated Annealing (SA) as a function of their cost are plotted. Points labeled (A)-(D) correspond to the energy minima $\mathscr E$ of respectively MS ($\mathscr E=-42\,465.3$), GA ($\mathscr E=-4\,154.1$), Hubs ($\mathscr E=-17\,626.2$) and SA ($\mathscr E=-36\,297.0$). Bottom and Left: number of activated nodes at each time step as a function of time for (A)-(D). The vertical scale for all plots (including the Top Right one) is identical.
}
\label{figEpinions}
\end{figure}

The figure also shows the distribution $P(t)$ of activation times, i.e. the number of nodes activated at a given time step of the dynamics. In the region where maximum spread is achieved, we observe bimodal activation time distributions for MS (e.g. at point (A)), for HUBS (e.g. at point (C)) and for SA (e.g. at point (D)): many nodes activate in the first steps of the dynamics, then the propagation proceeds at a slower pace, until a new acceleration finally induces a large fraction of the network to activate. The number of seeds found by MS at (A) is about $35$ times larger than the one found by Hubs at (C), but the corresponding cost is three times smaller: this implies that MS finds a larger set of scarcely influencing individuals that cooperatively are capable of initiating a large avalanche of activations. SA employs a cost per seed which is very close to the one of MS, but it generates a propagation extending to much longer times (which could be a drawback in applications). The existence of bootstrapping phenomena, responsible for the second peak in $P(t)$, also suggests that in this case GA may not be a well-suited optimization method, as the spread maximization problem does not satisfy submodularity and therefore the $63\%$-approximation bound to the size of the optimal cascade is not expected to hold.

\subsection{Role of Submodularity}

Most existing studies in this area are restricted to a class of propagation rules that obey a diminishing returns property called {\em submodularity}: the final (expected) spread size $f(A)$ as a function of the seed set $A$ is said to be submodular if $f(A\cup\{v\})-f(A) \geq f(B\cup\{v\})-f(B)$ for every $A\subseteq B$ and any vertex $v$.
This means that the differential gain obtained by adding any single vertex to a given set of seeds decreases (weakly) with the number of seeds already in the set, for any possible set of seeds.
Submodularity implies the absence of abrupt large-scale activation phenomena as a consequence of the addition of a small number of seeds.
 These cooperative effects  are instead widely observed in real model systems.
 In the model we shall consider (which is not submodular) the existence of cooperative effects are guaranteed by the fact that  $f(A\cup \{v\})-f(A)$ can increase abruptly with $|A|$. Existing algorithms based on greedy methods are known to perform very well on submodular systems \cite{KKT03,L07,CWW}, but obtain modest results on systems that are not submodular.
The relation of (the absence of) submodularity with the presence of cooperative effects can be seen as follows. Consider e.g. point (C) in Figure \ref{figEpinions} and the point just before the transition, i.e. before the addition of the last seed $v$. The difference in propagation is enormous, much bigger than what it would have obtained on earlier steps with the same seed (a similar transition is observed for the GA curve for a much larger number of seeds, outside of the plot to the right); this is exactly in contradiction with submodularity (i.e. the linear threshold model on this network is not submodular). On the other hand, the fact that such a large propagation was possible is due the the buildup of {\em cooperation} within the network, i.e. many nodes that get close to their threshold (favouring future reward) without crossing it (i.e. with no immediate ``self'' reward).

\section{Summary and conclusions}

We have put forward a method to solve inverse problems for irreversible dynamical processes defined over networks which is based on the cavity method of statistical physics and which is capable of identifying sets of seeds which maximize some objective function that depends on the activation times of each node.
Further progress can be achieved by extending the technique to stochastic irreversible dynamical models, such as the Independent Cascades Model, the LT model with stochastic threshold \cite{ABDZ} with arbitrary distributions, the Susceptible-Infected-Recovered model and or arbitrary signs in costs and revenues. This may be achieved by coupling the technique presented here with the stochastic optimization approach proposed in \cite{ABRZ11a, ABRZ11b}.

\begin{acknowledgments}
The authors acknowledge the european grants FET Open 265496, ERC 267915 and Italian FIRB Project RBFR10QUW4.
\end{acknowledgments}

\begin{appendix}

\section{Efficient update equations for BP}\label{ABP}

For the BP equations \eqref{update1bp}, we proceed in a way similar to that followed for MS. Here, however, the second constraint in \eqref{update1bp t positive} cannot be relaxed. Nonetheless, we notice that the sum on the right-hand side can be rewritten as
\begin{align}
 & \sum_{\substack{
	\{t_j, \, j \in \dd i \sm \ell\} \text{ s.t:} \\
	\sum_{k \in \dd i} w_{ki} \1 [t_k \leq t_i - 1] \geq \theta_i
  }}  \prod_{j \in \dd i \sm \ell} H_{ji}(t_j,t_i) \; -
 \sum_{\substack{
	\{t_j, \, j \in \dd i \sm \ell\} \text{ s.t:} \\
	\sum_{k \in \dd i} w_{ki} \1 [t_k \leq t_i - 1] \geq \theta_i \, ,\\
	\sum_{k \in \dd i} w_{ki} \1 [t_k < t_i - 1] \geq \theta_i
  }} \prod_{j \in \dd i \sm \ell} H_{ji}(t_j,t_i) \label{two_sums}\\
 = & \sum_{\substack{
	\{t_j, \, j \in \dd i \sm \ell\} \text{ s.t:} \\
	\sum_{k \in \dd i} w_{ki} \1 [t_k \leq t_i - 1] \geq \theta_i
  }}  \prod_{j \in \dd i \sm \ell} H_{ji}(t_j,t_i) \; -
 \sum_{\substack{
	\{t_j, \, j \in \dd i \sm \ell\} \text{ s.t:} \\
	\sum_{k \in \dd i} w_{ki} \1 [t_k < t_i - 1] \geq \theta_i
  }} \prod_{j \in \dd i \sm \ell} H_{ji}(t_j,t_i)
\end{align}
since for positive $w_{ki}$'s the second condition in the last sum of (\ref{two_sums}) implies the first one. We then introduce the following quantities
\begin{align}
  Q_{1, \dots, r}(\theta, t) & = \sum_{
	\substack{
	  \{ t_1, \dots, t_r \} \text{ s.t:} \\
	  \sum_{j = 1, \dots, r} w_j \1 \left[ t_j \leq t - 1 \right] = \theta
	}
  } \prod_{j = 1, \dots, r} H_{j}(t_j, t) \\
  Q'_{1, \dots, r}(\theta, t) & = \sum_{
	\substack{
	  \{ t_1, \dots, t_r \} \text{ s.t:} \\
	  \sum_{j = 1, \dots, r} w_j \1 \left[ t_j < t - 1 \right] = \theta
	}
  } \prod_{j = 1, \dots, r} H_{j}(t_j, t)
\label{convolutionbp Q}
\end{align}
where we have simplified the notation by relabeling the incoming messages and corresponding weights from 1 to $r$.
The single index $Q_j$ and $Q'_j$ are directly obtained from the definition:
\begin{align}
  Q_j(\theta, t) &= \sum_{
	\substack{
	  t_j \text{ s.t:} \\
	  w_j \1[t_j \leq t-1] = \theta
	}
  } H_j(t_j, t) \\
  &=
	\begin{dcases}
	  \sum_{t_j > t - 1} H_j(t_j, t) & \text{if $\theta = 0$,} \\
	  \sum_{t_j \leq t - 1} H_j(t_j, t) & \text{if $\theta = w_j$,} \\
      0 & \text{otherwise}
	\end{dcases} \label{convolutionbp Q1}
\end{align}
and similarly for $Q'_j$. The convolution of two $Q$ (or $Q'$) is given by
\begin{align}
	Q_{1, \dots, r}(\theta, t) &= \sum_{\theta', \theta'' \text{ s.t: } \theta' + \theta'' = \theta} Q_{1, \dots, s}(\theta', t) Q_{s+1, \dots, r}(\theta'', t)  \,.
\end{align}
Once $Q(\theta, t)$ and $Q'(\theta, t)$ are computed from the convolution of all the incoming edges, we can easily compute
\begin{align}
  H_{i\ell}(t_i, t_\ell) \propto & \sum_{\theta' \text{ s.t: } \theta' \geq \theta_i - w_\ell \1 \left[t_\ell \leq t_i -1 \right]} Q \big(\theta', t_i \big) - \sum_{\theta' \text{ s.t: } \theta' \geq \theta_i - w_\ell \1 \left[t_\ell < t_i -1 \right]} Q' \big(\theta', t_i \big) & \text{if $0 < t_i \leq T$} \,. \label{updatebp t positive}
\end{align}

Similarly, to compute \eqref{update1bp t d} we introduce
\begin{align}
  S_{1, \dots, r}(\theta) &= \sum_{
	\substack{
	  \{ t_1, \dots, t_r \} \text{ s.t: } \\
	  \sum_{j = 1, \dots, r} w_j \1 \left[ t_j < T \right] = \theta
	}
  }  \prod_{j = 1, \dots, r} H_j (t_j, \infty)
\label{convolutionbp S}
\end{align}
with single index $S$'s
\begin{align}
  S_j(\theta) &=
	\begin{dcases}
	   H_j(T, \infty) + H_j(\infty,\infty)  & \text{if $\theta = 0$} \\
	  \sum_{t_j < T} H_j(t_j, \infty) & \text{if $\theta = w_j$} \\
	  0 & \text{otherwise}
	\end{dcases} \label{convolutionbp S1}
\end{align}
and convolution
\begin{align}
  S_{1, \dots, r}(\theta) &= \sum_{\theta', \theta'' \text{ s.t: } \theta' + \theta'' = \theta}  S_{1, \dots, s}(\theta') S_{s+1, \dots, r}(\theta'') \,.
\end{align}
Once $S$ is computed for all the incoming edges, we can compute
\begin{align}
  H_{i \ell}(\infty, t_\ell) &\propto \sum_{\theta' \text{ s.t: } \theta' < \theta_i - w_{\ell i} \1 \left[ t_\ell < T \right]} S(\theta') e^{-\beta r_i}. \label{updatebp t d}
\end{align}

As in the case of MS, the update equations can be further simplified by noticing that in \eqref{update1bp} the dependence of $H_{i \ell}$ on $t_\ell$ is almost trivial: for fixed $t_i$, $H_{i \ell}$ is independent on $t_\ell$ if $t_i = 0$, it only depends on $\sign \big( t_\ell - (t_i - 1) \big) \in \{-1, 0, 1\}$ if $0 < t_i \leq T$, and it only depends on $\1 \left[ t_\ell < T \right]$ if $t_i = \infty$. We then introduce:
\begin{align}
  \tilde H_{i \ell}(t_i, \sigma) &=
	\begin{cases}
	  H_{i \ell}(t_i, t_i - 2) & \text{if $t_i > 1$ and $\sigma = 0$} \\
	  H_{i \ell}(t_i, t_i - 1) & \text{if $t_i > 0$ and $\sigma = 1$} \\
	  H_{i \ell}(t_i, t_i) & \text{if $\sigma = 2$} \\
	  0 & \text{otherwise}
	\end{cases} \label{H tildebp}
\end{align}
which can be inverted as
\begin{align}
  H_{i \ell}(t_i, t_\ell) &= \tilde H_{i \ell} \big( t_i, \hat \sigma(t_i, t_\ell) \big) \label{H vs H tildebp} \\
  \hat \sigma(t_i, t_\ell) &= 1 + \sign \big( t_\ell - (t_i - 1) \big) \label{sigma hatbp}
\end{align}
and in terms of which \eqref{convolutionbp Q1} and \eqref{convolutionbp S1} become respectively:
\begin{align}
  Q_j(\theta, t) &=
	\begin{dcases}
	  \sum_{t_j > t - 1} \tilde H_j \big( t_j, \hat \sigma(t_j, t) \big) & \text{if $\theta = 0$} \\
	  \sum_{t_j \leq t - 1} \tilde H_j(t_j, 2) & \text{if $\theta = w_j$} \\
	  0 & \text{otherwise}
	  \end{dcases} \\
  S_j(\theta) &=
	\begin{dcases}
	 \tilde H_j(T,2) + \tilde H_j(\infty, 2)  & \text{if $\theta = 0$} \\
	  \sum_{t_j < T} \tilde H_j( t_j, 2) & \text{if $\theta = w_j$} \\
	  0 & \text{otherwise.}
	\end{dcases}
\end{align}

\section{Alternative algorithms}\label{Aalgo}

\subsection{Linear Programming Algorithm}
We tried two different L/IP formulations: one closely related to our representation based on $x^t_i$ and the same dynamical constraints $\phi_i$ and a second one based on the completion of the partial ordering given by direct node activation to a total ordering \cite{ABW}, solving both with IBM's CPLEX software. In both cases, the computation time was very large, becoming unpractical for graphs of a few dozen nodes.

\subsection{Greedy Algorithms}
Greedy algorithms iteratively add to the seed set the vertex that maximizes a seed preference function. The latter
can be either the energy variation, or built upon some topological property like some centrality estimation. The
preference functions we considered were:
\begin{itemize}
\item GA: Largest marginal spread (in case of draw, choose e.g. the largest in-degree). This is computed by testing each
potential seed independently and integrating the dynamics.
\item HUBS: In-degree (a naive measure of influence).
\item HITS: Kleinberg's HITS centrality procedure. Centrality values have to be recomputed at each time step. HITS centrality is computed by the power method, so some quantities can be reused to make the computation more efficient. For HITS, weights and thresholds can be rescaled to make thresholds equal to 1 at every time step; this did not bring any improvement over the simpler version.
\end{itemize}
At each time step, the node with largest preference is chosen and added to the seed set and propagation is computed, then preferences are recomputed. To cope with seed costs, the preference function is complemented with a linear term including the seed prize. The linear coefficient was chosen to optimize the overall performance.

\subsection{Simulated Annealing}
One can use a simple Monte Carlo algorithm to bias the search in the configuration space of seed sets in order to minimize the energy function $\mathscr{E}$.
However, the activation-times representation is not the most convenient one to perform optimization by Monte Carlo, because of the presence of hard constraints.
On the other hand, such constraints are naturally taken into account if we consider the original dynamical formulation of the problem. At each Monte Carlo step, we have to compare the energies associated with two
configurations of seeds, the current one $\bx$ and the proposed one $\bx'$, that means one has to perform the deterministic dynamical process explicitly. With a naive implementation, a single propagation of the dynamical process requires $O(N T)$ operations, because of the parallel update of the nodes.
Hence, Monte Carlo methods have in this context an intrinsic limitation, due to the large number of operations that are necessary to perform in a single ``seed-change'' update on the configuration of seeds.
When $T$ is large it is much more convenient to adopt a different propagation algorithm, in which one keeps track only of the set $B_t$ of nodes that activate at time $t$. The set $B_0$ is composed of the set of seeds $S$. We also define a variable threshold value $\hat{\theta}_i^t$, such that $\hat{\theta}_i^0=0$ if $i\in S$ and $\hat{\theta}_i^0=\theta_i$ otherwise. We define $\partial B_t$ as the set of nodes $j$ such that $j \not\in B_t$ and $\exists i \in B_t$ for which the directed edge $(i,j) \in E$. For each node $j \in \partial B_t$, we define $n_j^t$ as the number of such links emerging from $B_t$ and connecting to $j$. The dynamics is defined simply by updating the threshold values of nodes $j \in \partial B_t$ as $\hat{\theta}_j^{t+1} = \hat{\theta}_j^t - n_j^t$ and $B_{t+1}$ is composed by all nodes $j \in \partial B_t$ such that $\hat{\theta}_j^{t+1}=0$. Using this algorithm, the computational time required for each propagation is at most $O(|E|)$, when the dynamics passes through all nodes within a time $T$. In summary, each Monte Carlo move requires $O(|E|)$ operations, making the computational time of the SA extraordinary demanding for large graphs.

We tried both linear and exponential annealing schedules, but they did not present relevant differences in the results; instead is some cases an accurate choice of the initial set of seeds turned out to improve considerably the performances of the SA. This is particularly important in real instances, where both the topology and the cost/revenue values are not uniform and possibly correlated (see discussion about SA in the Epinions network in Appendix \ref{Aepi}).

\section{Simulated Annealing on the Epinions graph}\label{Aepi}

\begin{figure}
\includegraphics[width=12cm]{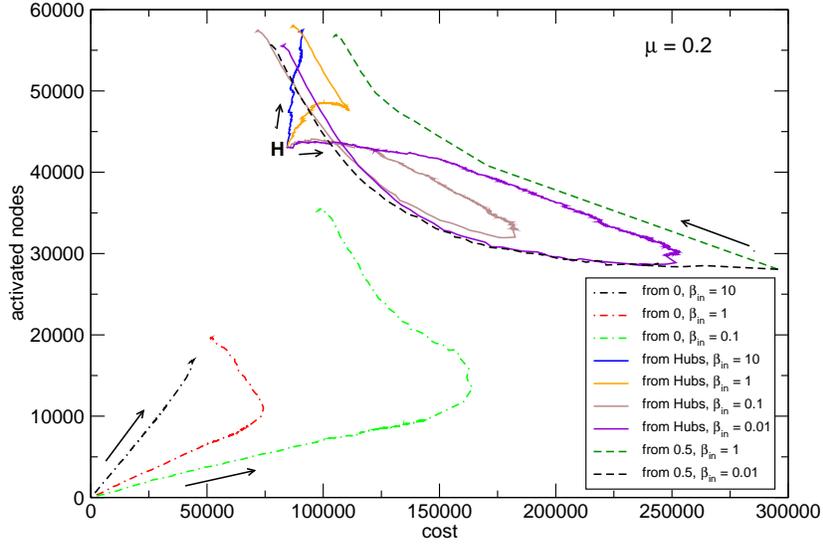}
\caption{Diagram of the number of activated nodes as function of the cost of the seeds obtained using SA on the Epinions graph with $\mu=0.2$, $\beta_{fin}=10^3$, and different values of $\beta_{in}$ as well as different initial configuration of seeds: configuration with zero seeds (dash-dotted lines), configuration with the $166$ most connected hubs (full lines)  and random initial conditions (dashed lines).
\label{fig:mc1}
}
\end{figure}

In the case of the Epinions graph ($N \simeq 76\,000$ and $|E| \simeq 500\,000$), the SA takes several hours to perform a single Monte Carlo sweep in the region of extensive propagations. In these conditions, a slow annealing schedule is practically impossible, therefore we limited our investigations to fast schedules reaching a final $\beta_{fin} = 10^3$ in $O(10^7)$ proposed moves of single seed change (corresponding to about one hundred of MC sweeps). In the following, we report some results for the case in which the costs of the seeds scale linearly with the out-degree of the nodes. In Figure \ref{fig:mc1} we considered a series of simulations with energy function $\mathscr{E}=\sum_i\{c_i\1[t_i=0]-r_i\1[t_i<\infty]\}$ where $c_i = \mu(k_i^{in}+1)+1$ and $r_i=1, \forall i$. We start from different types of initial conditions with $\mu =0.2$. The overall computational time is about one week on a single CPU.  Due to the limited number of MC steps, the performances SA strongly depend on the initial inverse temperature $\beta_{in}$. A fast cooling ($\beta_{in}=10$) generates a sort of steepest descent process that provides a larger number of finally activated nodes but with a larger total cost compared to the initial one. This is true for initial conditions with zero seeds (dash-dotted black line) as well as starting from the configuration of $166$ seeds found using the Hubs algorithm (dotted blue line). For smaller values of $\beta_{in}$, the SA is instead able to reach larger propagations by first adding many seeds and then slowly decreasing the overall cost. This is particularly notable when one starts from point H obtained using Hubs.  At the end of the process, the overall cost is lower than the initial cost due to the $166$ hubs contained in configuration corresponding to H. Decreasing $\beta_{in}$, the final results gets better (full lines starting from $H$). We compared these results with those obtained starting from completely random initial conditions at high temperature (dashed lines). In this case the lowest energy, representing the best tradeoff between cost and spread, is achieved for $\beta_{in}=0.1$ ($\mathscr{E} = -43\,066$). The results obtained using SA are reasonably good, although the computational complexity poses serious limitations on the length of the simulations and forces to use rather fast annealing schedules.

\begin{figure}
\includegraphics[width=12cm]{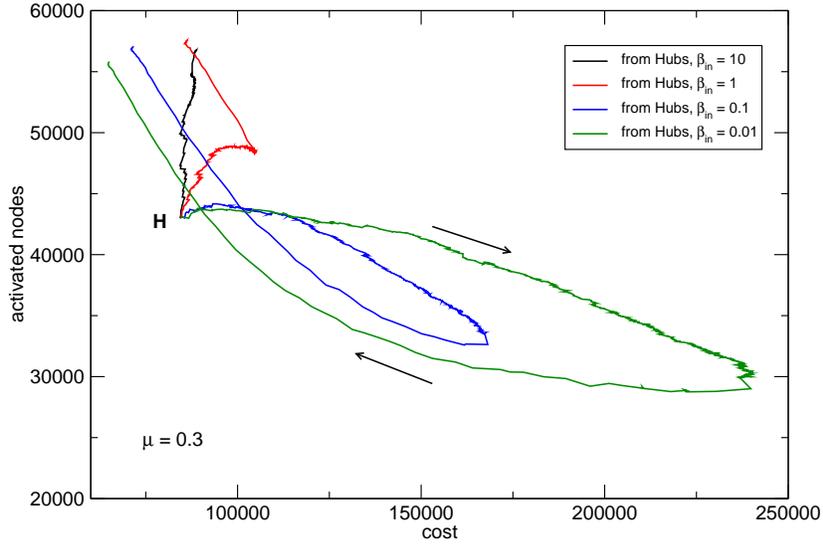}
\caption{Diagram of the number of activated nodes as function of the cost of the seeds obtained using SA on the Epinions graph with $\mu=0.3$, $\beta_{fin}=10^3$, and different values of $\beta_{in}$ starting from a configuration containing the $166$ most connected hubs as seeds.
\label{fig:mc2}
}
\end{figure}

For $\mu = 0.3$, we can directly compare the results of SA with those of the MS algorithm reported in the main paper. Figure \ref{fig:mc2} displays only the results obtained starting from point $H$ (i.e. $166$ seeds selected using the Hubs heuristic) and different $\beta_{in}$. The lowest energy value achieved in the simulations is $\mathscr{E} = -36\,297$ as compared to the best result obtained using the Max-Sum algorithm, $\mathscr{E} = -42\,465$.

\begin{figure}
\includegraphics[width=12cm]{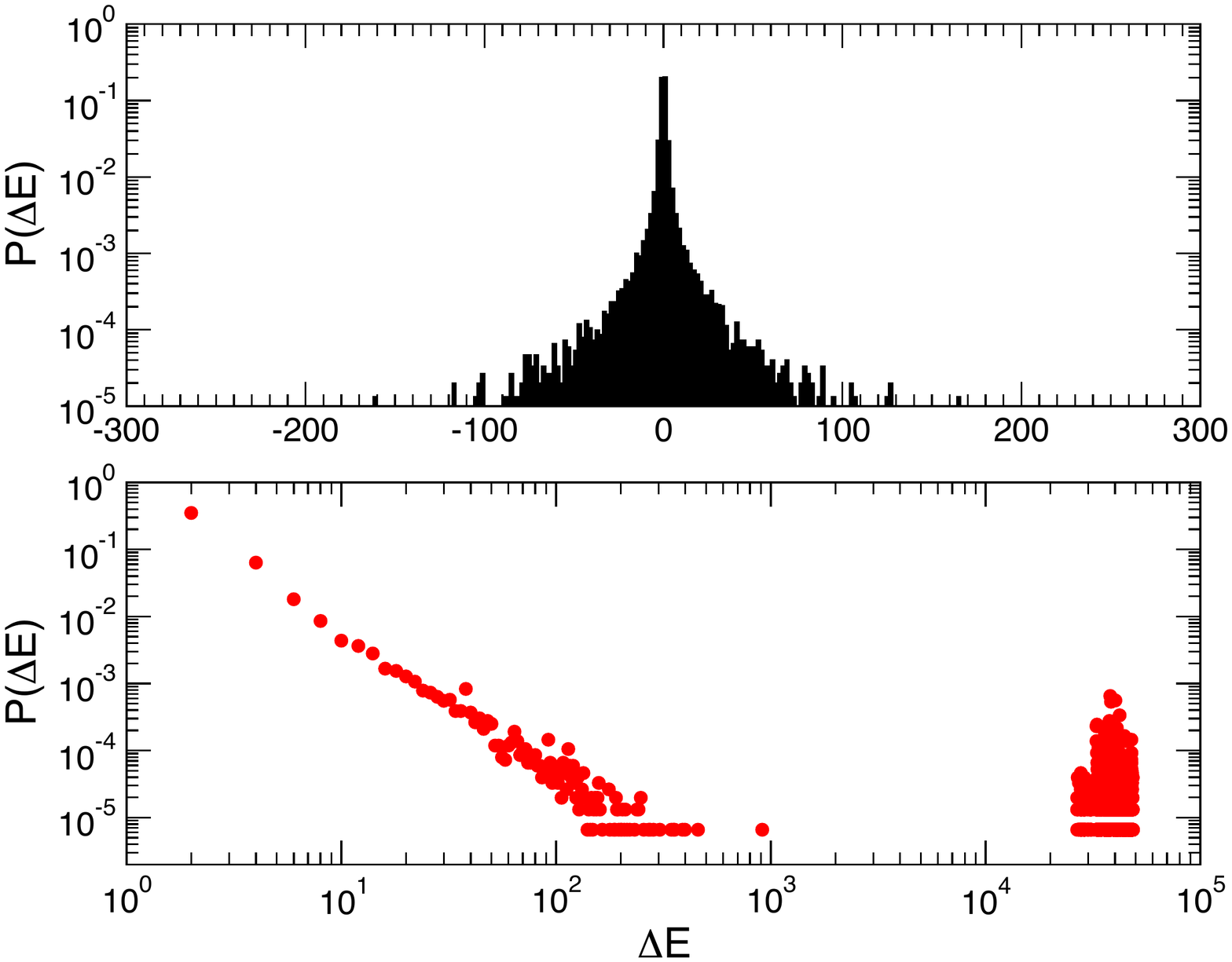}
\caption{Histogram of the energy variations $\Delta\mathscr{E}$ associated to single seed changes, computed on two configurations of seeds: Top) a randomly generated configuration in which each node is chosen as seed with probability 0.5 and Bottom) a local minimum reached after a SA simulation with $\mu=0.3$ and $\beta_{in}=0.1$.
\label{fig:mc3}
}
\end{figure}

It is interesting to notice that at low temperatures the SA gets stuck into local minima of the energy landscape. This is clearly visible from the distribution of the energy variations $\Delta \mathscr{E}$ corresponding to possible changes of a single seed in different states. Figure \ref{fig:mc3}A corresponds to a completely random configuration of seeds; panel B refers instead to local minima reached starting from point $H$. It is not unlike to find very large values of $\Delta \mathscr{E}$ and, in some cases, the distribution $P(\Delta \mathscr{E})$ is non zero over almost five order of magnitudes of energies. It means that at the same temperature, different moves can have extremely different acceptance rate, making the optimization process much more complex at low temperatures.

\end{appendix}


\begin{thebibliography}{XX}

\bibitem{CLR79}
Chalupa, J., Leath, P. L. \& Reich, G. R.  (1979) Bootstrap percolation on a Bethe lattice. {\em J. Phys. C: Solid State Phys.} {\bf 12} L31.

\bibitem{G78}
Granovetter, M. (1978) Threshold Models of Collective Behavior. {\em Am. J. Sociol.} {\bf 83}, 1420-1443.

\bibitem{R83}
Rogers, E. M. (1983) {\em Diffusion of Innovations,} New York: Free Press Rogers, .

\bibitem{JY05}
Jackson, M. \& Yariv, L.  (2005) Diffusion on social networks. {\em Economie Publique} {\bf 16}, 69-82.

\bibitem{AOY11}
Acemoglu, D., Ozdaglar, A., \& Yildiz, M.E. (2011) Diffusion of Innovations in Social Networks, {\em Proc. of IEEE Conference on Decision and Control}.

\bibitem{NE02}
Newman, M. E. J. (2002) Spread of epidemic disease on networks. {\em Phys. Rev. E} {\bf 66} 016128.

\bibitem{W02}
Watts, D. J. (2002) A Simple Model of Global Cascades on Random Networks. {\em PNAS} {\bf 99}, 5766-5771.

\bibitem{OCK13}
O'Dea,~R., Crofts,~J. J. and Kaiser,~M. (2013) {\em J. R. Soc. Interface} {\bf 10}(81), 1742-5662

\bibitem{financial-risk}
Eisenberg,~L. and Noe,~T. H. (2001) Systemic Risk in Financial Systems, {\em Management Science} {\bf  47}, 236-249.

\bibitem{Nier2007}
Nier,~E., Yang,~J., Yorulmazer,~T. and Alentorn,~A. (2007) Network models and financial stability, {\em Journal of Economic Dynamics and Control} {\bf 31} 2033-2060.

\bibitem{Gai2010}
Gai,~P. and Kapadia,~S. (2010) Contagion in financial networks, {\em Proceedings of the Royal Society} {\bf A 466} (2120) 2401-2423.

\bibitem{HM11}
Haldane, A. G. \& May, R. M.  (2011) Systemic risk in banking ecosystems. {\em Nature} {\bf 469}, 351Ð355.

\bibitem{BBV08}
A. Barrat, M. Barth\'elemy, and A. Vespignani, {\em Dynamical Processes on Complex Networks},
Cambridge University Press, Cambridge (2008)

\bibitem{K10}
Kitsak, M. et al.  (2010)  Identification of influential spreaders in complex networks. {\em Nature Physics} {\bf 6}, 888-893.

\bibitem{nostro}
Altarelli,~F., Braunstein,~A., Dall'Asta,~L., and Zecchina,~R., (2013) accepted for publication, {\em Phys. Rev. E}.

\bibitem{DR01}
Domingos, P., and Richardson, M. Mining the network value of customers,in {\em KDD'01} (2001)

\bibitem{KKT03}
Kempe, D., Kleinberg, J. \& Tardos, E.  (2003) Maximizing the Spread of Influence through a Social Network.
{\em Proc. 9th ACM SIGKDD Intl. Conf. on Knowledge Discovery and Data Mining}.

\bibitem{CLRS01}
Cormen, T. H.,  Leiserson, C.E.,  Rivest, R.L., and Stein. C. {\em Introduction to Algorithms}, Second Edition. MIT Press and McGraw-Hill (2001)

\bibitem{L07}
Leskovec, J., Krause, A., Guestrin, C., Faloutsos, C., Van-
Briesen, J., and Glance, N.S. Cost-effective outbreak detection in networks, in {\em KDD'07} (2007).

\bibitem{CWW}
Chen, W., Wang, C., and Wang, Y. Scalable influence maximization for
prevalent viral marketing in large-scale social networks, in {\em ACM SIGKDD Int. Conf. on Knowledge Discovery and Data Mining} (2010).
Chen, W., Wang, Y., and Yang, S. Efficient influence maximization in
social networks, in {\em KDD'09} (2009).
Chen, W., Yuan, Y., and Zhang, L. Scalable influence maximization in
social networks under the linear threshold model, in {\em ICDM'10} (2010).

\bibitem{K07}
Kleinberg, J.  (2007) Cascading Behavior in Networks: Algorithmic and Economic Issues. In {\em Algorithmic Game Theory} (N. Nisan, T. Roughgarden, E. Tardos, V. Vazirani, eds.), Cambridge University Press.

\bibitem{mining}
Goyal, A., Bonchi, F., and Lakshmanan, L.V. A data-based approach to
social influence maximization, in {\em PVLDB'12}  (2012).

\bibitem{LZWKF}
Lu, Z., Zhang, W., Wu, W., Kim, J. \& Fu, B.  (2011) The complexity of influence maximization problem in the deterministic linear threshold model.
{\em J. Comb. Optim.} {\bf 24-3} 374-378.


\bibitem{NWF78}
Nemhauser, G.,  Wolsey, L., and Fisher, M. An analysis of the approximations
for maximizing submodular set functions. {\em Mathematical Programming}, {\bf 14}:265Ð294 (1978)


\bibitem{MP}
M\'ezard, M. and Parisi, G. (2001) The Bethe lattice spin glass revisited, {\em Eur. Phys. J.} B {\bf 20}, 217.


\bibitem{MM09}
M\'ezard, M. and Montanari, A. (2009) Information, Physics and Computation, Oxford graduate texts.

\bibitem{steiner}
M. Bayati, C. Borgs, A. Braunstein, J. Chayes, A. Ramezanpour, R. Zecchina. (2008) Statistical mechanics of Steiner trees {\em Physical Rev. Lett.} {\bf 101} (3), 37208;
M. Bailly-Bechet, C. Borgs, A. Braunstein, J. Chayes, A. Dagkessamanskaia, J. M. Francois, R. Zecchina (2011) Finding undetected protein associations in cell signaling by belief propagation. {\em Proc. Nat. Acad. Sci} {\bf 108} (2), 882-887;
I. Biazzo, A. Braunstein, R. Zecchina (2012) Performance of a cavity-method-based algorithm for the prize-collecting Steiner tree problem on graphs. {\em Physical Review E} {\bf 86} (2), 026706


\bibitem{epinions}
Dataset from J. Leskovec's SNAP collection at http://snap.stanford.edu/data/.

\bibitem{K99}
Kleinberg, Jon M. (1999) Authoritative sources in a hyperlinked environment. {\em J. ACM}, 46(5):604-632.

\bibitem{ABDZ}
Altarelli, F., Braunstein, A., Dall'Asta, L. and Zecchina, R., In preparation.

\bibitem{ABRZ11a}
Altarelli, F., Braunstein, A., Ramezanpour, A., and Zecchina, R., (2011) Stochastic optimization by message passing, {\em J. Stat. Mech.} P11009.

\bibitem{ABRZ11b}
Altarelli, F., Braunstein, A., Ramezanpour, A., and Zecchina, R., (2011) Stochastic Matching Problem, {\em Phys. Rev. Lett.} {\bf 106} 190601.


\bibitem{ABW}
Ackerman, E., Ben-Zwi, O. \& Wolfovitz, G.  (2010) Combinatorial model and bounds for target set selection. {\em Theor. Comp. Sci.} {\bf 411}, 44-46.

\end{thebibliography}
\end{document}